\documentclass[12pt]{article}
\usepackage{amsmath,amsfonts,amssymb,latexsym}
\usepackage{epsfig}
\usepackage{graphicx}

\def\D{\mathbb{D}}
\def\S{\mathbb{S}}

\begin{document}

\begin{center}
\bf{\LARGE
Operator algebra in the space of images}\footnote{Contribution to the 8th International Workshop DICE2016, Sept.12-16, 2016, \\
Castiglioncello (Italy). Submitted to  {\it J. Phys: Conf. Series.}}
\end{center}
\bigskip\bigskip

\begin{center}
E. Celeghini
\end{center}

\begin{center}
{\sl Dipartimento di Fisica, Universit\`a  di Firenze and
INFN--Sezione di
Firenze \\
I50019 Sesto Fiorentino,  Firenze, Italy\\
\medskip
Departamento de Fisica T\'eorica, Universidad de Valladolid, E-47005, 
Valladolid, Spain}
\\
{e-mail: celeghini@fi.infn.it}
\end{center}

\bigskip
\bigskip

\begin{abstract}

A consistent description of images on the disk and of their transformations
is given as elements of a vector space and of an operators algebra.
The vector space of images on the disk $\D$ is the Hilbert space $L^2(\D)$
that has as a basis the Zernike functions.
To construct the operator algebra that transforms the images,\, $L^2(\D)$ must be complemented and the full rigged Hilbert space $RHS(\D)$  considered. Only 
this rigged Hilbert space
allows indeed to write the operators of different cardinality we need to build  the ladder operators on the Zernike functions that by inspection, belong to the representation 
$D_{1/2}^+ \otimes D_{1/2}^+$ of the algebra $su(1,1) \oplus su(1,1)$.
Consequently the transformations of images  are operators contained inside the universal
enveloping algebra $UEA[su(1,1) \oplus su(1,1)]$. 
Because of limited precision of experimental measures, physical states can be always described by vectors of the Schwartz space $\S(\D)$, dense in the $L^2(\D)$ space 
where the manipulation of images is performed.

\end{abstract}

\vspace{1cm}

\section{Introduction}

Images are essential in our life and in our research.
The problem is that the information contained in an image is
too large in order that the human mind can manage it and
we have to isolate the limited relevant information. 
To our opinion, this work  to clean the signal from spurious or forgettable elements
and highlight the interesting information could be improved  
combining computer action and  
human ability of synthesis. 

The only approach in this elaboration of images 
that has obtained up to now relevant results 
is the cleaning-up realized by adaptive optics where an auxiliary photoreceptor 
measures the wave front deformations introduced by the medium 
and acts on the instrument to induce the opposite effect. 
Adaptive optics removes indeed the spurious phases in the complex function 
$f(r,\theta)$ that represents the optical signal on the disk allowing to obtain the cleaned 
image $|f(r,\theta)|^2$.

We are proposing here that $|f(r,\theta)|^2$
 can be consider not necessarily the final result of the process but 
possibly an intermediate step that 
can be further elaborated by means of an action we can call
soft adaptive optics.

While hard adaptive optics acts only on the perturbations of the phase 
introduced by the medium, this soft elaboration of the numerical image 
$|f(r,\theta)|^2$ operates on a set of pixels to obtain another set of pixels,
independently from the cause of the distortion (wind 
in atmosphere, diffraction in lents, defects of the apparatus ...) and the particular instrument
of measure.

More, while hard adaptive optics can be applied only to the cleaning of images, the 
proposed soft approach converts images into images and can be employed
everywhere a transformation of images can play a role, like -for instance- laser physics, microscopic images, radioastronomy or in general instrumental improvement.

The construction of a theory of operators that act in the space of images defined on the disk, 
transforming images into images, is the subject of this paper. 
The action on images/functions defined on a rectangle is similar and will be
discussed elsewhere \cite{Ce2}.
\vspace{.2cm}

 Characteristics of soft adaptive optics are: 
 \begin{enumerate}
 \vspace{-.2cm}
 \item It can be applied both\, on\, and\,  of-line,\, so it can be used later with particular
~~accuracy when convenient.
\vspace{-.1cm}
\item Consuming only time machine, it has not mechanical moving parts and
\vspace{-.1cm}
\item it allows to consider together images from different sources like optical and 
radio images.
\end{enumerate}
\vspace{.1cm}

 The relevant mathematical points are:
 \vspace{-.2cm}
\begin{itemize}
 \item
Images on the disk belong to the Hilbert space $L^2(\D)$\,  with basis the Zernike 
functions \cite{BoWo}.
\vspace{-.1cm}
\item
To introduce together continuous and discrete operators we need to construct the ladder operators on the Zernike functions we have to complement $L^2(\D)$ considering the
rigged Hilbert space on the disk $(RHS(\D))$ i.e. \{$\S(\D) \subset L^2(\D) \subset \S(\D)^x\}$.
\vspace{-.1cm}
\item
The Hilbert space $L^2(\D)$\, is isomorphic to the unitary irreducible representation
$D_{1/2}^+ \otimes D_{1/2}^+$\;\, of the Lie group\, $SU(1,1) \otimes SU(1,1)$\,.
\vspace{-.1cm}
\item
$L^2(\D)$\, operators belong to the Universal Enveloping Algebra (UEA)
$UEA[su(1,1)\oplus su(1,1)]$\, and
\vspace{-.1cm}
\item
can be computed algebraically or by means of first  order differential operators.
\end{itemize}

The mathematical theory of rigged Hilbert spaces can be found in \cite{Bo, ReSi, GaGo}.
A physically relevant -basis dependent- approach is discussed in \cite{Ce15, CeGaOl, CeGaOl2}, centered on the two points:
\begin{enumerate}
\vspace{-.2cm}
\item In physics every measured quantity has a limited precision so that it can be 
described by an element of the Schwartz space $\S(\D)$ dense into the 
Hilbert space $L^2(\D)$.
\vspace{-.1cm}
\item The inclusion map moves basis vectors from the dual of $\S(\D)$\, -$\S(\D)^x$-\, 
into the dual of $L^2(\D)$, $L^2(\D)^x \equiv L^2(\D)$. 
\end{enumerate}

In sect.2 we introduce the Zernike functions. In sect.3 we 
sketch the functional analysis of $RHS(\D)$. In sect.4 we 
introduce the operators, their algebraic structure and 
their realization. In sect.5 we give a description of the practical modus operandi
in images manipulation. 

\section{Zernike functions}

Let us start from  the radial Zernike polynomials on the circle, {$R^m_n(r)$}\cite{BoWo},
where\, {$n$}\, is a 
natural number and\, {$m$}\, an integer, such that\, {$0\leq m\leq n$}\, with\, {$n-m$}\, even.
$\{R^m_n(r)\}$\; are real polynomials defined for\, 
{$0\leq r \leq 1$}
such that  $R^m_n(1) = 1$  and
are usually given in explicit form or in terms of hypergeometric functions. 
We prefer to define them by means of their second order  differential equation, fundamental in
our discussion:
\[
 \left[(1-r^2)\frac{d^2}{dr^2} -(3 r-\frac{1}{r}) \frac{d}{dr} + n(n+2) -\frac{m^2}{r^2}\right]\,
 R^m_n(r) =0.
\]  
For\, {$m$}\, fix,\, $\{R_n^m(r)\}$\, satisfy
\[\begin{array}{lll}\displaystyle
\int_0^1 \;R^{{m}}_n(r)\; R^{{m}}_{n'}(r)\;r\; dr &=& \displaystyle \frac{\delta_{n\, n'}}{~2(n+1)~}\;,
\\[0.4cm] \displaystyle
\sum_{n=m}^\infty \; (n+1)\; R_n^{{m}}(r)\, R_n^{{m}}(r') &=&  
\displaystyle\frac{\delta(r-r')}{~2 r~} \,.
\end{array}\]
As the interest in optics is focused on real functions defined on the disk,
starting from $R_n^m(r)$ it is usual to introduce the functions
\[
{\cal Z}^{-m}_n(r,\theta) := R^m_n(r) \;{\rm sin}(m \theta)\qquad {\cal Z}^{m}_n(r,\theta) := 
R^m_n(r) \;{\rm cos}(m \theta)
\]
and also to resume the two indices $n$ and $m$ in a unique sequential
index \cite{No}.

However this paper is centered on symmetry and
the objects "invariant in form" respect to rotations \cite{BhWo}
are those defined in the complex space by Born and Wolf \cite{BoWo}:
\[
Z_n^m(r,\theta) := R_n^{|m|}(r)\; e^{{\bf i} m \theta}
\]
with $n$ and $m$ integer, $n-m$ even, but $-n \leq m \leq n$.
The symmetry can be further improved  writing $n$ and $m$ in function of 
two natural numbers $k$ and $l$ \cite{Du}
\[
n = k+l\,\qquad m = k-l
\]
and introducing a multiplicative factor. The final objects we consider here are thus 
\[
V_{k,l}(r,\theta)\; := \sqrt{k+l+1}\;\; R_{k+l}^{|k-l|}(r)\; e^{{\bf i}(k-l)\theta}
\qquad\qquad ( k, l = 0, 1, \dots )
\]
that have the symmetries
\vspace{-.2cm}
\[
V_{l,k}(r, \theta) = {V_{k,l}(r, \theta)}^* = V_{k,l}(r, -\theta) 
\]
\vspace{-.2cm}
and satisfy
\begin{equation}\begin{array}{l}\label{3} \displaystyle
\frac{d^2 V_{k,l}(r, \theta)}{dr^2} = \\[0.3cm] \displaystyle
\quad \frac{1}{1-r^2} \left[\left(3 r-\frac{1}{r}\right) \frac{d}{dr}
-(k+l)(k+l+2)+\frac{(k-l)^2}{r^2} \right] V_{k,l}(r, \theta)
\end{array} \end{equation}
 They are orthonormal on the unit disk\, $\D$\, :
\begin{equation}\label{1}
\int^{2 \pi}_0 d \theta \int^1_0 dr\; r\; {V_{k,l}(r,\theta)}\; V_{k',l'}(r,\theta)^*\, 
=\, \delta_{k,k'}\, \delta_{l,l'}
\end{equation}
so that they are a basis in the Hilbert space $L^2(\D)$ of complex square
 integrable functions defined on the unit disk $\D$.

They also satisfy
\begin{equation}\label{2}
\sum_{k,l\,=0}^\infty  {V_{k,l}(r,\theta)}\; V_{k,l}(r',\theta')^*\, =\, \frac{1}{r}\, \delta(r-r')\, 
\delta(\theta - \theta') 
\end{equation}
exhibiting that the space of images is not a simple
Hilbert space but of a more complex structure, a RHS.
As explained in in sect.3 and, in more detail, in \cite{CeGaOl2}, the sum (\ref{2}) is indeed 
outside $L^2(\D)$: in fact it converges to a tempered distribution in the topology of 
$\S(\D)^x$, a space contained in $RHS(\D)$ but larger of $L^2(\D)$. 
However this and the related convergence problems are irrelevant in applications
where, because of experimental errors, only elements of $\S(\D) \subset L^2(\D)$ are considered. 

As the set $\{V_{k,l}(r,\theta)\}$\, is a basis in $L^2(\D)$,
every function  $f(r,\theta) \in L^2(\D)$  can be written
\[
f(r,\theta) = \sum_{k,l}\, f_{k,l}\;\, V_{k,l}(r,\theta)
\]
\vspace{-.2cm}
where \cite{ReSi}
\vspace{-.2cm}
\begin{equation}\label{4}
f_{k,l} := \frac{1}{\pi}\, \int_0^{2 \pi} d\theta \int_0^1 dr\; r\;\, 
f(r,\theta)\,  V_{k,l}(r,\theta)^*\;\; 
\qquad {\rm with}\:\: \sum_{k,l} | f_{k,l}|^2  < \infty .
\end{equation}
The completeness determines the inner product and the Parceval identity (\ref{5})
\[
\frac{1}{\pi}\, \int_0^{2 \pi} d\theta \int_0^1 dr\, r\;   f(r,\theta)\;  g(r,\theta)^*\;
=\;  \sum_{k,l}\; f_{k,l}\; {g_{k,l}}^*\; ,
\]
\begin{equation}\label{5}
\frac{1}{\pi}\, \int_0^{2 \pi} d\theta \int_0^1 dr\, r\;\,  |f(r,\theta)|^2\,  =\,  
\sum_{k,l}\, |f_{k,l}|^{~2} .
\end{equation}

Now we can introduce in\, $\{V_{k,l}(r,\theta)\}$\, the operators\, $R$,\, $D_R$,\, 
$\Theta$, $D_\Theta$, $K$ and $L$ such that:
\[
~~~~{R}~\, V_{k,l}(r,\theta)\,: =\,  {r}\, V_{k,l}(r,\theta)~,~~~~~
~~{D_R}~\, V_{k,l}(r,\theta)\,:=\, \frac{{d} V_{k,l}(r,\theta)}{{dr}}~,
\]
\[
~~~~~~~~{\Theta}~\, V_{k,l}(r,\theta)\,:=\, {\theta}\, V_{k,l}(r,\theta)~,~~~~~
~~{D_\Theta}~\, V_{k,l}(r,\theta)\,:=\,  
\frac{{d} V_{k,l}(r,\theta)}{{d\theta}}~~,~~~
\]
\[
 ~~~{K}~\, V_{k,l}(r,\theta)\,:=\, {k}\, V_{k,l}(r,\theta)~,~~~~~~~~
{L}~\, V_{k,l}(r,\theta)~\,:=\, {l}~\, V_{k,l}(r,\theta)~,
\]
\noindent that allow to write eq.(\ref{3}) as an operatorial identity  on the whole space 
$\{V_{k,l}(r,\theta)\}$
\begin{equation}\label{6}
D^{~2}_R\,= \frac{1}{1-R^2}\left[(3 R-\frac{1}{R}) D_R -(K+L)(K+L+2)
+\frac{1}{R^2}(K-L)^2 \right]
\end{equation}
i.e.  for each  $f(r,\theta) \in \{V_{k,l}(r,\theta)\}$ the operator $D^{~2}_R$  is not independent 
but related to a first order differential operator. 

\section{Functional analysis of functions on the disk}

The general discussion of $RHS(\D)$ will be presented in \cite{CeGaOl2}. 
We limit ourselves here to sketch the fundamentals statements in the Zernike basis.

The Rigged Hilbert space on the disk, $RHS(\D)$, is the Gel'fand triple \cite{ReSi}
\[     
\S(\D) \;\subset\; L^2(\D) \;\subset\; \S(\D)^x \,.
\]
$\S(\D)$\, is a Schwartz\; space,  sub-space of $L^1(\D)$, defined as
\[
\S(\D) :=\;\; \{ \,f(r,\theta) \in C^\infty(\D)\,\;  |\;\;  \Arrowvert f(r,\theta)
\Arrowvert_{\alpha, \beta, \gamma, \delta} \,<\, \infty 
\quad\;\, \forall\, \alpha,\,\beta,\,\gamma,\,\delta <\, \infty \}
\]
\[
{\Arrowvert f(r,\theta) \Arrowvert}_{\alpha, \beta, \gamma, \delta} \;:=\;\;\;    
{sup}_{\{0 \leq r<1,\, 0 \leq \theta<2\pi\}}\, \;\;|\, r^\alpha\, \theta^\beta \,{D_r}^\gamma\, {D_\theta}^\delta f(r,\theta)\; | \;.
\]
$\S(\D)^x$\, (that includes the temperated distributions) is the dual 
of\, $\S(\D)$\,, i.e. the space of the applications from  $\S(\D)$\, into  the field of complex numbers\, ${\bf C}$ 
\[
\langle \,\S(\D)^x\, |\, \S(\D)\, \rangle \,\in\,  {\bf C}.
\]
$L^2(\D)$ is a Hilbert space. It is the completion of $\S(\D)$\, i.e. it contains every Cauchy
sequence of\, $\S(\D)$ and $\S(\D)$\, is dense in\, $L^2(\D)$\,. 

The\; {Riesz representation theorem}\; states that\,
$L^2(\D)$\, is isomorphic to its dual $L^2(\D)^x$.
$f(r,\theta) \in L^2(\D)$ if eq.(\ref{4}) is satisfied and $f(r,\theta) \in \S(\D)$\, if \cite{ReSi}
\begin{equation}\label{7}
\sum_{k,l}\, (k+1)^2 (l+1)^2\, |f_{k,l}|^2\, <\, \infty
\end{equation}

The inclusion map allows to move\, $\langle V_{k,l}(r,\theta)|$\,  from\, $\S(\D)^x$\, to\, 
$L^2(\D)^x \equiv L^2(\D)$. So, for all\,  $V_{k,l}(r,\theta)$,
\[
\langle V_{k,l}(r,\theta)|V_{k',l'}(r',\theta')\rangle_{{\S(\D)^x \otimes \,\S(\D)}}  \;\;\;\;=\;\;\;\;  
\langle V_{k,l}(r,\theta)|V_{k',l'}(r',\theta')\rangle_{L^2(\D)^x  \otimes  \,L^2(\D)}
\]
\noindent i.e.,  $RHS(\D) \equiv L^2(\D)$ for $f(r,\theta) \in \S(\D)$.
This is exactly the situation we have in applications where, as every 
measure is subject to an error of measure, every experimental $f(r,\theta)$ can 
be considered to fulfill  eq.(\ref{7}) and thus such that $f(r,\theta) \in \S(\D)$.
So that for a physicist the only difference between a RHS and a Hilbert
space is that the RHS allows to consider together operators of different
cardinality ($R$, $\Theta$, $D_R$ and $D_\Theta$ from one side and $K$ and $L$ from the other) while Hilbert space not.

\section{Operators and algebra}

The fundamental benefit of $RHS(\D)$ compared to $L^2(\D)$ is indeed 
that in $RHS(\D)$ we can give a rigorous description of  all (discrete and continuous) 
operators we need to realize the algebra of rising and lowering operators. 

Let us indeed introduce in\, $\{V_{k,l}(r,\theta)\}$ the recurrence operators:
\begin{equation}\begin{array}{l}\label{8}
A_+\, V_{k,l}(r, \theta)\;=\; (k+1)\, ~V_{k+1,l}(r, \theta) ,
\\[.3cm]
A_-\, V_{k,l}(r, \theta)\;=\; k\, ~V_{k-1,l}(r, \theta) , 
\\[.3cm]
B_+\, V_{k,l}(r, \theta)\;=\; (l+1)\, ~V_{k,l+1}(r, \theta) ,
\\[.3cm]
B_-\, V_{k,l}(r, \theta)\;=\; l\, ~V_{k,l-1}(r, \theta)\; .
\end{array}\end{equation}
They, by inspection, are functions of\, $R$, $D_R$, $\Theta$, $K$ and $L$ : 
\[
 ~~~~A_+ :=~~ \frac{e^{+ {\bf i} \Theta}}{2} \; \left[-(1-R^2) D_R + R (K+L+2) +\frac{1}{R}(K-L)\right]
 \sqrt{\frac{K+L+2}{K+L+1}} ,
 \]
 \[
  A_- :=~~ \frac{e^{ -{\bf i} \Theta}}{2} \; \left[+(1-R^2) D_R + R (K+L) +\frac{1}{R}(K-L)\right]~ \sqrt{\frac{K+L}{K+L+1}} ,
 \]
 \[
 ~~~~B_+:=~~ \frac{ e^{- {\bf i} \Theta}}{2} \; \left[-(1-R^2) D_R + R (K+L+2) -\frac{1}{R}(K-L)\right]
 \sqrt{\frac{K+L+2}{K+L+1}} ,
 \]
 \[
 ~B_-:=~~ \frac{ e^{+ {\bf i} \Theta}}{2} \; \left[+(1-R^2) D_R + R (K+L) -\frac{1}{R}(K-L)\right]~\sqrt{\frac{K+L}{K+L+1}}\; .
\]

The discrete operators $K$ and $L$ are diagonal on\, $\{V_{k,l}(r,\theta)\}$ but, 
as the parameters $k$\; and 
\;$l$ are modified by the action of $A_\pm$ and $B_\pm$,
do not commute with $A_\pm$ and $B_\pm$:
\begin{equation}\label{9}
[ K, A_\pm] = \pm A_\pm\;, \qquad [ L, B_\pm ] = \pm B_\pm \,,
\end{equation}
exhibiting that not only eq.(\ref{2}) but also the operator structure needs  the $RHS(\D)$.

Now as\, $A_\pm$\,, $B_\pm$\,, $K$ and $L$ are operators, we can apply them iteratively
in\, $\{V_{k,l}(r,\theta)\}$\, and, in particular, we can calculate their commutators 
\[
[A_+,A_-]\; V_{k,l}(r,\theta)= -2(k+1/2)\; V_{k,l}(r,\theta)\,, \;\,\,\,\quad [K,A_\pm]\, V_{k,l}(r,\theta)= 
\pm \,V_{k \pm 1,l}(r,\theta)\,.
\]
So, defining\;\, $A_3 := K+1/2$\,,\; we see that\, $\{A_+,\, A_3,\, A_-\}$\, are on\; 
$\{V_{k,l}(r,\theta)\}$ a differential realization of the
Lie algebra\; $su(1,1)$:
\[
[A_+,\, A_-]\,=\,-2 A_3\;, \qquad [A_3,\, A_\pm]\,=\,\pm A_\pm \,
\]
and, analogously, as
\[
[B_+,B_-]\; V_{k,l}(r,\theta)= -2(l+1/2)\; V_{k,l}(r,\theta) \,,\,\,\quad [L,B_\pm] V_{k,l}(r,\theta)=
 \pm \,V_{k,l \pm 1}(r,\theta)\, ,
\]
$\{B_+,\; B_3:= L+1/2\; ,\, B_-\}$\, are on\; $\{V_{k,l}(r,\theta)\}$
one other algebra\; $su(1,1)$
\[
[B_+,\, B_-]\,=\,-2 B_3 \qquad [B_3,\, B_\pm]\,=\,\pm B_\pm \,.
\]
Finally,  as $A_i$ and $B_j$ commute in\; $\{V_{k,l}(r,\theta)\}$, we can complete the algebra with 
\[
[A_i, B_j] = 0 ,
\] 
defining thus a  differential realization of the 6 dimensional Lie algebra\; $su(1,1)\oplus su(1,1)$ .

The explicit calculation of the $su(1,1)$ Casimir invariants gives
\[
C_A\;  V_{k,l}(r,\theta)\,=\,\left[\frac{1}{2} \{ A_+, A_- \} - A_3^2\right]\, V_{k,l}(r,\theta)\, =\, \frac{1}{4}\, 
V_{k,l}(r,\theta) ,
\]
\[
C_B\;  V_{k,l}(r,\theta)\,=\,\left[\frac{1}{2} \{ B_+, B_- \} - B_3^2\right]\, V_{k,l}(r,\theta)\, =\, \frac{1}{4}\, 
V_{k,l}(r,\theta) .
\]
As the Casimir of the discrete series\, $D^+_j$  of\, $su(1,1)$\, is\; $j(1-j)$\, with 
$j=1/2, 1,..$ \cite{Ba}, $\{V_{k,l}(r,\theta)\}$\, is isomorphic
to the representation\; $D^+_{1/2} \otimes D^+_{1/2}$ 
of the group\; $SU(1,1)\otimes SU(1,1)$. 

Now we can move from the algebra\; $su(1,1)\oplus su(1,1)$\; to the associated universal 
enveloping algebra (UEA) i.e. the algebra\;   
UEA[$su(1,1)\oplus su(1,1)$]  that has, as a basis, the ordered monomials
\;$A_+^{\alpha_1} A_3^{\alpha_2} A_-^{\alpha_3} B_+^{\beta_1} B_3^{\beta_2} B_-^{\beta_3}$ 
(where\; $\alpha_i$ and ${ \beta_j}$\,  are natural
numbers) submitted to the \,$su(1,1)\oplus su(1,1)$\, commutation relations.
Every operator ${\cal O} \in$ UEA[$su(1,1)\oplus su(1,1)$]\; can be written 
\begin{equation}\label{10}
{\cal O} = \sum_{\bar{\alpha}, \bar{\beta}} {\cal O}_{\bar{\alpha}, \bar{\beta}} = 
\sum_{\bar{\alpha}, \bar{\beta}} c_{\bar{\alpha}, \bar{\beta}} \; A_+^{\alpha_1} A_3^{\alpha_2} 
A_-^{\alpha_3} B_+^{\beta_1} B_3^{\beta_2} B_-^{\beta_3} ,
\end{equation}
where $c_{\bar{\alpha}, \bar{\beta}}$ are constants depending from ${\bar{\alpha}}$ and 
${\bar{\beta}}$. 

Of course, as\, $\{V_{k,l}(r,\theta)\}$\, is a differential representation of the algebra\; $su(1,1)\oplus su(1,1)$,
it is also a differential representation of\; $UEA[su(1,1)\oplus su(1,1)]$.

Because the representation $D^+_{1/2} \otimes D^+_{1/2}$ is unitary and 
irreducible the vector space
of unitary operators acting on the space $L^2(\D)$ is isomorphic to the set of operators acting on
$D^+_{1/2} \otimes D^+_{1/2}$
i.e. all invertible transformations of one image in another image
can be written in the form (\ref{10}), belong to
$UEA[su(1,1)\oplus su(1,1)]$ and are differential operators in $L^2(\D)$.

${\cal O}$, in principle, can be a differential operator of higher order but, by 
means of iterated use of eqs.(\ref{6}, \ref{9}), can be always reduced in the space\,
$\{V_{k,l}(r,\theta)\}$ 
to an equivalent first order differential operators.
Let us stress that in mathematics the UEA contains the closure of polynomials
(i.e. the series) that in physical applications, because of the measure errors, can be excluded.

Thus, the operators that relate two arbitrary images are polynomials
of the UEA and can be computed algebraically by means of iterated applications of
eqs.(\ref{8}) or analytically
as first order differential operators.

\section{Applications to images}

Starting from $f(r,\theta)\,\in L^2(\D)$,  
images on the disk\, $\D$\, are described as\, $|f(r,\theta)|^2$
that,  disregarding the phases, we relate to $|f(r,\theta)|\,\in L^2(\D)$. By means of 
eq.(\ref{4}) we write its components in terms of basic vectors:
\[
f_{k,l} = \frac{1}{\pi}\, \int_0^{2 \pi} d\theta \int_0^1 dr\; r\;\, |f(r,\theta)|\, V_{k,l}(r,\theta)^*\;\; 
\]
where, as $|f(r,\theta)|$ is real, $f_{l,k} = (f_{k,l})^*$.

In applications we are limited to values of $k$ and $l$ such that the Parceval identity, 
eq.(\ref{5}), is satisfied in the approximation consistent with the experimental errors in\, $|f(r,\theta)|$,
writing the finite sum:
\begin{equation}\label{11}
|f(r,\theta)| \approx \sum_{k,l=0}^{k_M,l_M}  f_{k,l}\; V_{k,l}(r,\theta) 
\end{equation}
where, as $R_{k+l}^{|k-l|}(r)
$ are polynomials, $k_M$ and $l_M$ are quite smaller 
if\, $|f(r,\theta)|$ is a smooth function.

Now, as discussed before, any operator ${\cal O}$ that perform the 
transformation from the starting image  $|f(r,\theta)|$ 
to a new arbitrary chosen image $|g(r,\theta)|$ can be written in the form eq.(\ref{10})
with finite sums in $\bar{\alpha}$ and $\bar{\beta}$.

At this point we have two computational options. Following the algebraic approach, we can continue writing
\[
{\cal O}_{\bar{\alpha}, \bar{\beta}}\; |f(r,\theta)|\; =\;  \sum_{k,l}^{k_M, l_M}\;\, f_{kl} \;
 c_{\bar{\alpha}, \bar{\beta}}\;\; A_+^{\alpha_1} A_3^{\alpha_2} 
A_-^{\alpha_3} B_+^{\beta_1} B_3^{\beta_2} B_-^{\beta_3}\; V_{k,l}(r,\theta)
\]
and, with iterated applications of eqs.(\ref{8}),
calculating
\[
A_+^{\alpha_1} A_3^{\alpha_2} 
A_-^{\alpha_3} B_+^{\beta_1} B_3^{\beta_2} B_-^{\beta_3}\; V_{k,l}(r,\theta) .
\]
We obtain, in this way, the coefficients\, $g_{k,l}$ that satisfy
\[
A_+^{\alpha_1} A_3^{\alpha_2} 
A_-^{\alpha_3} B_+^{\beta_1} B_3^{\beta_2} B_-^{\beta_3}\; V_{k,l}(r,\theta)\; =\;
g_{k+\alpha_1-\alpha_3, l+\beta_1-\beta_3}\; V_{k+\alpha_1-\alpha_3,\, l+\beta_1-\beta_3}(r,\theta) 
\]
and we have
\[
{\cal O}_{\bar{\alpha}, \bar{\beta}}\; |f(r,\theta)|\, =\,  \sum_{k,l}\; f_{kl} \;\,
 c_{\bar{\alpha}, \bar{\beta}}\;\; g_{k+\alpha_1-\alpha_3,\, l+\beta_1-\beta_3}\; 
 V_{k+\alpha_1-\alpha_3,\, l+\beta_1-\beta_3}(r,\theta)\; ,
\]
that allows to write our final result:
\[
{\cal O}\; |f(r,\theta)| = |g(r,\theta)|\;.
\]

In the alternative analytical approach, the operator ${\cal O}$ can be
described in terms of it differential expression: combining in ${\cal O}$ iterated applications
of eq.(\ref{6}) and eq.(\ref{9}) all dependence from the operators $K$ and $L$
can be moved to the right and all power of $D_R$ higher of one reduces to one,
obtaining 
\begin{equation}\label{12}
{\cal O}\, =\, \sum_i\;\, \left[f_i(R) D_R  +   h_i(R)\right]\, g_i(K,L)\; .  
\end{equation}
Applying this first order differential operator ${\cal O}$ -as written in eq.(\ref{12})- 
to $\sum f_{k,l}\; V_{k,l}(r,\theta)$ -as defined in eq.(\ref{11})- we have again:
\[
|g(r,\theta)| = \sum_{k,l}\, f_{k,l}\;\, {\cal O}\;{V_{k,l}(r,\theta)} .
\]

\vspace{1cm}


\end{document}